
\input harvmac

\input epsf
\def\caption#1{{\it
	\centerline{\vbox{\baselineskip=12pt
	\vskip.1in\hsize=5.0in\noindent{#1}\vskip.1in }}}}
\def\pyidk{PHY-9057135}
\def\frac#1#2{{\textstyle{#1\over #2}}}
\def\ltap{\ \raise.3ex\hbox{$<$\kern-.75em\lower1ex\hbox{$\sim$}}\ }
\def\gtap{\ \raise.3ex\hbox{$>$\kern-.75em\lower1ex\hbox{$\sim$}}\ }
\ifx\epsffile\undefined\message{(FIGURES WILL BE IGNORED)}
\def\insertfig#1{}
\else\message{(FIGURES WILL BE INCLUDED)}
\def\insertfig#1{{{
\midinsert\centerline{\epsfxsize=\hsize
\epsffile{#1}}\bigskip\bigskip\bigskip\bigskip\endinsert}}}
\fi
\def\CM{\cal M}
\def\bfm#1{\rlap{$#1$}\mkern-0.75mu #1}
\def\vec#1{\bfm{#1}}

\def\veci#1#2{{\vec #1}_{#2}}
\def\pin{{\vec p}_{\rm in}}
\def\pout{{\vec p}_{\rm out}}

\def\bra#1{\left\langle #1 \right|}
\def\ket#1{\left| #1 \right\rangle}

\def\CS{{\cal S}}
\def\({\left(}
\def\){\right)}
\def\[{\left[}
\def\]{\right]}
\def\np#1#2#3{Nucl. Phys. B{#1} (#2) #3}
\def\pl#1#2#3{Phys. Lett. {#1}B (#2) #3}
\def\prl#1#2#3{Phys. Rev. Lett. {#1} (#2) #3}
\def\physrev#1#2#3{Phys. Rev. {#1} (#2) #3}

\Title{
\vbox{
\hbox{DOE/ER/40561-302-INT96-00-156 }
\hbox{UW/PT-96-32}
\hbox{UCSD/PTH 96-10}}}
{The Nucleon-Nucleon Potential in the $1/N_c$ Expansion}


\centerline{David B. Kaplan${}^a$ and Aneesh V.~Manohar${}^b$}
\bigskip
\centerline{${}^a${\sl Institute for Nuclear Theory, University of
Washington,}}
\centerline{{\sl Box 351550, Seattle, WA 98195-1550}}
\centerline{${}^b${\sl Department of Physics, University of California at San
Diego,}}
\centerline{{\sl La Jolla, CA 92093-0319}}


\vskip .3in
The nucleon-nucleon potential is analysed using the $1/N_c$ expansion of QCD.
The $NN$ potential is shown to have an expansion in $1/N_c^2$, and the
strengths
of the leading order  central, spin-orbit, tensor, and quadratic spin-orbit
forces (including isospin dependence) are determined. Comparison with a
successful phenomenological potential (Nijmegen) shows that the large-$N_c$
analysis explains many of the qualitative features observed in the
nucleon-nucleon interaction. The $1/N_c$ expansion implies an
effective Wigner supermultiplet symmetry for light nuclei.
Results for baryons containing strange quarks are presented in an
appendix.

\Date{12/96} 

\newsec{Introduction}

The two-body nucleon-nucleon interaction is the basic ingredient that is used
in calculating the properties of nuclei. There are a number of
phenomenologically successful models for the interaction, typically constructed
using  meson exchange contributions.  Unfortunately there is no direct
connection between these models and the underlying  theory of the strong
interactions, QCD.  Until recently, the only way that QCD and the physics of
nucleon interactions could be rigorously related has been through symmetry
arguments.  By making use of symmetry and effective field theory, one can
calculate the low energy dynamics using a small number of parameters that are
fitted to the experimental data. Such theories can be quite predictive, but
still remain somewhat remote from QCD --- one would like {\it a priori}
arguments for the sizes of the phenomenological parameters.

Recently, there has been significant progress in better understanding the
implications of QCD for hadronic physics by exploiting the ``hidden'' expansion
parameter of QCD --- $1/N_c$, where $N_c=3$ is the number of colors
\nref\dm{R.~Dashen and A.V.~Manohar, \pl {315} {1993} {425,
438}.}\nref\ej{E.~Jenkins, \pl {315} {1993} {431, 441,
447}.}\nref\djm{R.~Dashen, E.~Jenkins and A.V.~Manohar, \physrev
{D49}{1994}{4713}, \physrev {D51} {1995}{3697}.}\nref\cgo{C.~Carone, H.~Georgi,
and  S.~Osofsky, \pl {322} {1994} {227}.}\nref\lmr{M.~Luty and
J.~March-Russell, \np {426} {71} {1994}.}\refs{\dm-\lmr}. In the large $N_c$
limit, one finds that meson-baryon interactions respect an $SU(4)$ spin-flavor
symmetry --- the same  symmetry found in the nonrelativistic quark model quark
model  \nref\gs{J.-L.~Gervais and B.~Sakita, \prl{52} {1984} {527}, \physrev
{30} {1984} {1795}.}\refs{\dm,\gs}. It has also been shown that the $1/N_c$
expansion can provide information about the nucleon-nucleon
potential~\ref\ks{D.B. Kaplan  and M.J. Savage,  \pl {365}{1996}{244}.}.  In
particular, ref.~\ks\ analyzed the central potential for $NN$ scattering and
showed that the $1/N_c$ expansion gave a qualitative understanding of its spin
and isospin structure.

In this paper we extend the analysis of \ks\ to include the entire $NN$
potential. Naively, one might think that the $N_c\rightarrow \infty$ limit is
not relevant for analyzing nuclear physics. Nuclear matter forms a classical
crystal at $N_c=\infty$, and so there must be a phase transition between
$N_c=3$ and $N_c=\infty$.  While
the $1/N_c$ expansion is not reliable for studying bulk properties of nuclear
matter, it does allow one to analyze the  spin and isospin dependence of the
nuclear force. One expects that the symmetry properties of the $NN$ interaction
will be independent of the phase of the many body groundstate.

The general form of the potential for elastic, nonrelativistic $NN$
scattering is
\eqn\genpot{\eqalign{
V_{NN} &= V^0_0 + V^0_\sigma \sigma_1\cdot\sigma_2 + V^0_{LS} {\bf L}\cdot{\bf
S} +
V^0_T S_{12} + V^0_Q Q_{12}\cr
&+ \(V^1_0 + V^1_\sigma \sigma_1\cdot\sigma_2 + V^1_{LS} {\bf
L}\cdot{\bf S} + V^1_T S_{12} + V^1_Q Q_{12}\)\tau_1\cdot \tau_2
\ ,}}
where
\eqn\tensors{\eqalign{
S_{12}&\equiv 3 {\bf \sigma}_1\cdot{\bf \hat r}\ {\bf \sigma}_2\cdot{\bf \hat
r} - {\bf \sigma}_1\cdot{\bf \sigma}_2 \cr
Q_{12} &= \half\left\{({\bf \sigma}_1\cdot{\bf L}),({\bf \sigma}_2\cdot{\bf
L})\right\}\ .}}
The four terms $V^i_0$, $V^i_\sigma$ constitute the central potential, while
$V^i_T$, $V^i_{LS}$ and $V^i_Q$ are the tensor interaction, the spin-orbit
interaction,  and the quadratic spin-orbit interaction respectively; the ten
functions $V^i_a$ can in general be velocity dependent. The main result of this
paper is that the strength of  the ten functions  $V^i_a$ can be determined in
the $1/N_c$ expansion, and are  as given  in Table~1. As is apparent from this
Table, the actual expansion parameter is not $1/N_c$ but $1/N_c^2$.  Thus even
though the actual value $N_c=3$ is not very large, an  expansion in $1/N_c^2$
can be quite predictive.
\topinsert
\centerline{TABLE 1}
\centerline{\vbox{
\medskip
\tabskip=0pt \offinterlineskip
\def\space#1{height #1 pt&\omit&&\omit&&\omit&&\omit&&\omit&&\omit&\cr}
\def\space#1{height #1 pt&\omit&&\omit&&\omit&&\omit&&\omit&&\omit&\cr}
\halign{\vrule # &\quad\hfil # \hfil\quad&& \vrule # &
\quad\hfil $ # $ \hfil\quad \cr
\noalign{\hrule}
\space 3
& {\rm Isospin }  && V_0 && V_\sigma && V_{LS} && V_T
 && V_Q &\cr
\space 3
\noalign{\hrule}
\space 3
& ${\bf 1\cdot 1}$ && N_c && 1/N_c && 1/N_c && 1/N_c && 1/N_c^3 &\cr
\space 3
\space 3
&${\bf \tau_1\cdot \tau_2}$ &&1/N_c && N_c && 1/N_c && N_c && 1/N_c &\cr
\space 3
\noalign{\hrule}
}}
}
\bigskip
\endinsert

The organization of this paper is as follows: in \S2 we briefly review general
properties of baryons in the large $N_c$ limit.  In \S3 we derive the results
given in Table~1. These results are compared in \S4 with the ``Nijmegen
potential'' of references \nref\niji{M.M. Nagels, T.A. Rijken, J.J. de Swart,
\physrev{D17}{1978}{768}}\nref\nijii{V.G.J. Stoks, R.A.M. Klomp, C.P.F.
Terheggen, J.J. de Swart, \physrev{C49}{1994}{2950}}\refs{\niji,\nijii}  --- a
phenomenologically successful model of the $NN$ interaction; we show that the
the hierarchy of Table~1 is evident in $NN$ phenomenology.  \S5 extends  the
discussion of ref.~\ks\ concerning how the Wigner supermultiplet symmetry might
arise in light nuclei as a consequence of the $1/N_c$ expansion.  This is
followed by conclusions, and an appendix in which we  extend our
analysis to hyperon interactions ({\it i.e}, including the $s$ quark).

\newsec{The Large $N_c$ QCD Analysis}

The large $N_c$ limit is defined by taking the number of colors $N_c$ of QCD to
be large while simultaneously rescaling the QCD coupling as $g\to
g/\sqrt{N_c}$, keeping $\Lambda_{QCD}$ fixed \nref\thooft{G. 't Hooft, \np
{72}{1974}{461}.}\nref\witten{E. Witten, \np{160}{1979}{57}.}\thooft. The
$1/N_c$ expansion has proven to be a powerful tool for analyzing baryon
properties, since baryon structure simplifies considerably in this limit.
Antiquarks in the baryon are   suppressed, and  as baryons consist of $N_c$
quarks interacting with $1/N_c$ strength,  the Hartree approximation becomes
exact in the large $N_c$ limit \witten.  Although one cannot solve the Hartree
equations due to the nonlinearity of glue interactions, one can nevertheless
determine a number of useful  properties of the spin and flavor properties of
the baryons and their interactions.

To analyze the flavor and spin structure of the Hartree Hamiltonian in the case
of two light flavors, it is convenient to use as an operator basis the
one-quark operators of the quark model
\eqn\quarkbasis{
\hat S^i = q^\dagger {\sigma^i\over 2} q,\qquad \hat I^a =
q^\dagger {\tau^a \over 2 } q,\qquad \hat
G^{ia} = q^\dagger {\sigma^i \tau^a \over 4} q,
}
where $q=(u,d)$ and $q^\dagger$ are the creation and annihilation operators for
the $u$ and $d$ quark flavors, and $\sigma^i, \tau^a$ are the standard $SU(2)$
Pauli matrices acting on spin and isospin respectively.  The $q$ and
$q^\dagger$ operators do not carry color, and are bosonic.  The Hartree
Hamiltonian can then be constructed as monomials of these operators. An
important result from large $N_c$ QCD is that the Hartree Hamiltonian takes the
form \refs{\djm-\lmr}:
\eqn\hartree{ H = N_c \sum_n \sum_{s,t} v_{stn} \( {\hat S\over N_c}\)^s\(
{\hat
I\over N_c}\)^t\( {\hat G\over N_c}\)^{n-s-t} }
where the operators $\{\hat S,\,\hat I,\,\hat G\}$ are given in
eq.~\quarkbasis, the coefficients $v$ are $\CO(1)$ functions of momenta, and we
have suppressed isospin, spin, and vector indices which are contracted such
that $H$ is rotation and isospin invariant. An example of a contribution to $H$
is pictured  in \fig\figi{ A QCD contribution to $H$ leading in
$1/N_c$.  This diagram can be described in spin-flavor space
as a product of three  of the 1-quark operators given in
eq.~\quarkbasis.}. It is important  that although we make use of the
quark model operator basis, eq.~\hartree\ makes no assumption about the
validity of the quark model; the quark model operators are a
representation of the spin-flavor Clebsch-Gordon coefficients,
and provide an efficient way of doing group theory
computations. A Skyrme model basis, for example, would have
worked just as well~\ref\am{A.V. Manohar, \np
{248}{1984}{19}.}.

\topinsert
\centerline{\epsfxsize=1.5in \epsfbox{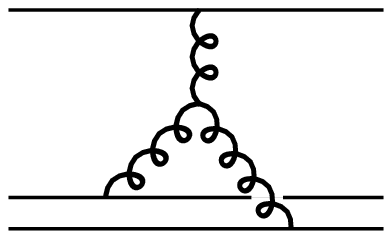}}
\caption{Fig.~1. A QCD contribution to $H$ leading in
$1/N_c$.  This diagram can be described in spin-flavor space
as a product of three  of the 1-quark operators given in
eq.~\quarkbasis.}
\endinsert

The lowest lying  eigenstates of $H$ with baryon number $B=1$ have isospin $I$
and spin $S$ satisfying  $I=S=\half, \frac{3}{2},\frac{5}{2}\ldots$.   The
first two states can be identified with the $N$ and $\Delta$;  the higher spin
states do not exist for $N_c=3$.  To leading order, this baryon tower is
degenerate with mass $M\sim N_c$.  One can show that it transforms as an
irreducible representation (the totally symmetric $N_c$-index tensor) of an
approximate $SU(4)$ spin-flavor symmetry;  this is the symmetry under which the
quark operators $u\uparrow$, $u\downarrow$, $d\uparrow$, $d\downarrow$
transform as the four-dimensional fundamental
representation~\nref\gs{J.-L.~Gervais and B.~Sakita, \prl{52} {1984} {527},
\physrev {30} {1984} {1795}.}\refs{\dm,\gs}.  For $N_c=3$ the $\{N,\Delta\}$
spin states transform as the 20 dimensional representation of $SU(4)$, familiar
from the quark model.

Additional information can be obtained by considering matrix elements of
operators between  baryon states $B$ and $B'$ restricted to have $S=I\sim 1$.
For example, matrix elements of  the basis operators \quarkbasis\ satisfy
\eqn\nsize{ \bra{B'}  {\hat S/ N_c}\ket{B} \sim \bra{B'}  {\hat I/ N_c}\ket{B}
\sim  1/N_c\ ,\qquad \bra{B'}  {\hat G/ N_c}\ket{B}\sim  1 \ .} Matrix elements
of many body operators can be analyzed as well, using various relations among
powers of the basic operators $\hat S^i$,  $\hat I^a$, and $ \hat G^{ia}$
\refs{\dm,\djm}. This allows one to greatly reduce the number of linearly
independent terms in \hartree\ at a given order in $1/N_c$.  Using these
techniques, one is able to show that the matrix element of a general  $n$-quark
operator $\hat\CO^{(n)}_{I,S}$ with  $B=0$, isospin $I$ and spin $S$, is of
size \djm\ (see also \ks\ for a derivation)
\eqn\ijrule{\bra{B'}
\hat\CO^{(n)}_{I,S}/N_c^n \ket{B} \ltap  1/N_c^{\vert I-S\vert}\ .}
The fact
that the operators with the largest matrix elements have $I=S$ was first
observed in the Skyrme model \ref\mattis{M.P. Mattis and M. Mukherjee, \prl
{61}{1988}{1344}\semi M.P. Mattis, \physrev {D39}{1989} {994}.}, and is known
as the ``$I_t=J_t$ rule''.

Eqs. \hartree --\ijrule\ are the central results behind the large $N_c$
analysis of baryons.  One consequence is that the mass splittings in the baryon
tower (e.g, between $N$ and $\Delta$) are of size $1/N_c$ \refs{\ej,\witten}.
This result is in good agreement with the real world where the ratio $R\equiv
(M_\Delta-M_N)/(M_\Delta + M_N)=0.13$, while the large $N_c$ prediction at
$N_c=3$ is $R\sim 1/N_c^2 =0.11$. Consequences of eq.~\nsize\ for the $NN$
interaction are explored in the next section.

\newsec{The Nucleon-Nucleon Interaction}

There are two independent three-momenta for baryon-baryon scattering in the
center of mass frame, which can conveniently taken to be
\eqn\qkdef{\vec q = \veci p {in} - \veci p {out}\ ,\qquad \vec k = \veci p {in}
+ \veci p {out}\ .}
These momenta are to be considered independent of $N_c$ in the $1/N_c$
expansion.
To leading order in $1/N_c$, the entire baryon tower is degenerate and $\vert
\veci p{in}\vert = \vert\veci p{out}\vert$ for elastic scattering, up to
$\CO(1/N_c^2)$ corrections,
and so $\vec q\cdot \vec k =0$ to the same order. The general baryon-baryon
interaction potential is then a matrix
\eqn\potdef{V(\vec q,\vec k)  = \bra{\veci p {out},\gamma; -\veci
p{out},\delta}
H\ket{\veci p{in},\alpha; -\veci p{in},\beta}}
where $H$ is the Hartree Hamiltonian \hartree\ and $\alpha,\ldots,\delta$
denote internal quantum numbers of the baryons, such as spin, flavor and
particle type ({\it e.g,} $N$ or $\Delta$). Throughout this paper we will
define the $NN$ potential as the above matrix element restricted to the space
of nucleons. We do not consider second order effects due to virtual
$\Delta$'s, etc.

There are two ways that $1/N_c$ factors can suppress terms in the potential.
The first arises from spin-flavor structure and the powers of $1/N_c$ in
eq.~\nsize. The second source of suppression arises in velocity dependent
interactions arising as relativistic corrections. Since the nucleon velocity
equals $\vec p/M\sim 1/N_c$, each power of velocity is equivalent to a $1/N_c$
suppression. In the nonrelativistic limit for baryons, a $t$-channel meson
exchange
contribution to $V$ is only a function of $\vec q$. A $u$-channel contribution
is only a function of $\vec k$, and can be expressed as an exchange potential.
Relativistic corrections allow a single meson exchange contribution to $V$ to
be a function of both $\vec q$ and $\vec k$. Meson exchange in the $t$-channel
is then a function of $\vec q$ and $\vec k/M$, with each power of $\vec k$
being accompanied by one factor of $M$. Similarly, $u$-channel meson exchange
is a function of $\vec k$ and $\vec q/M$. This shows that if a general velocity
dependent potential is expanded in a Taylor series in $\vec k$ and $\vec q$, a
term of the form $\vec q^r\vec k^s$ is suppressed by
\eqn\kqsupp{1/N_c^n\ ,\qquad n={\rm Min}(r,s)\ .}
Combining this source of $1/N_c$ suppression with eq.~\nsize\ will allow us to
determine the  size and spin-flavor structure of the dominant terms in the
potential $V$.

An $NN$ interaction at the QCD level gets contributions from complicated
processes, such as pictured in  \fig\figiii{An example of a contribution to the
$NN$ interaction at the level of quarks and gluons.  This diagram can be
described in spin-flavor space as a single 1-quark operator acting on the first
baryon $N_1$, and two 1-quark operators acting on the $N_2$ line. Nothing
physical depends on how one assigns the final quark lines to $N_1$ or $N_2$, so
long as one considers all possible interactions.}. Each of these contributions
can be expressed as a tensor function $v({\vec q},{\vec k})$ contracted with
1-quark operators $\hat S^i$, $\hat I^a$ and $\hat G^{ia}$ which act on either
of the two nucleon states.  The coefficient function $v_{stn}$ in eq. \hartree\
and the  operators $\hat S^i$ and
$\hat G^{ia}$ must combine to be invariant under rotations.  Our analysis is
simplified  by first expanding  $v_{stn}$ (and hence $V$) in multipole moments;
 the
following subsections are organized accordingly as $\Delta L=0$ (the central
force), $\Delta L=1$ (spin-orbit force), and $\Delta L=2$ (the tensor and
quadratic spin-orbit forces).
\topinsert
\centerline{\epsfxsize=2.0in \epsfbox{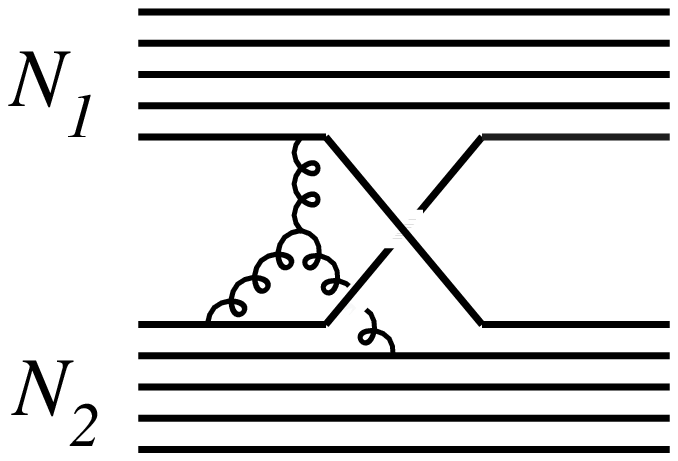}}
\caption{Fig.~2. An example of a contribution to the $NN$ interaction from the
3-quark operator pictured in fig.~1.  This diagram can be described in
spin-flavor space as a single 1-quark operator acting on the first baryon
$N_1$, and two 1-quark operators acting on the $N_2$ line.}
\endinsert

\subsec{$\Delta L=0$: The Central Potential}

This is the case analyzed in ref. \ks. The central force can be  written as a
sum of products of 1-quark operators as in
eq.~\hartree, where the operators act on either the $N_1$ or $N_2$ nucleon
states, and the coefficients $v_{stn}$ are general scalar functions of
$\vert{\vec q}\vert$ and $\vert{\vec k}\vert$.  It follows from eq.~\nsize\
that the leading contribution will have no $\hat S^i/N_c$ or $\hat I^a/N_c$
operators, since each of these implies a $1/N_c$ suppression;  instead it will
consist solely of powers of $\hat G^{ia}/N_c$.  By rotational symmetry, since
the $v$ coefficients are scalars, the $\hat G^{ia}$ operators must be
contracted to form spin invariants. Similarly, isospin symmetry implies that
the $\hat G^{ia}$ must be contracted to form isospin invariants.  From these
constraints, it is possible to show that at leading order in $1/N_c$, the most
general form for the central potential is~\ks
\eqn\vcentral{
V_{\rm central} = N_c\sum_{n=0}^{N_c} v_n
\left({\hat G_1\cdot \hat G_2\over N_c^2}\right)^n,
}
where $\hat G_1\cdot \hat G_2 \equiv \hat G_1^{ia} \hat G_2^{ia}$. In general,
the coefficients $v_n$ are functions of both $\vert\vec q\vert^2$, $\vert\vec
k\vert^2$, and obey the rule eq.~\kqsupp. One can further restrict the powers
of $\hat G_1\cdot \hat G_2$ in eq.~\vcentral\ to be completely symmetric in the
$\hat G_1$ indices, and in the $\hat G_2$ indices, before the two sets of
indices are contracted.

It is straightforward to verify that \vcentral\ is the most general form of the
leading order $\Delta L=0$ potential. We have argued that it can only involve
powers of the $\hat G^{ia}$ operators, on the basis of eq.~\nsize;  what must
be shown is that the indices are contracted as above in eq.~\vcentral. By the
operator reduction rule \djm\ any terms in which two indices of $\hat G^{ia}
\hat G^{jb}$ (where both $\hat G$'s act on the same baryon) are contracted with
each other by $\delta^{ij}$, $\delta^{ab}$, $\epsilon^{ijk}$ or
$\epsilon^{abc}$ can be eliminated in favor of terms with fewer powers of $\hat
G$ Thus the only allowed invariants are obtained by contracting the indices
of $\hat G_1^{ia}$ with those of $\hat G_2^{ia}$, as in \vcentral. More
complicated contractions, such as
\eqn\compi{
\hat G_1^{ia}\ \hat G_1^{jb}\ \hat G_2^{ib}\ \hat G_2^{ja}
}
can be written as
\eqn\comp{\eqalign{
&\left(\hat G_1\cdot \hat G_2\right)^2  +
\hat G_1^{ia}\ \hat G_1^{jb} \left[
\hat G_2^{ib}\ \hat G_2^{ja}- \hat G_2^{ia}\ \hat G_2^{jb}\right]\cr
=&\left(\hat G_1\cdot \hat G_2\right)^2  - \epsilon^{abc}\,
\hat G_1^{ia}\, \hat G_1^{jb} \ \epsilon^{ghc}\,
\hat G_2^{ig}\, \hat G_2^{jh}.
}}
The term with two $\epsilon$ symbols can be reduced to $\hat G_1\cdot \hat G_2$
using the relations in \djm, so that all contractions of $\hat G_1$ with $\hat
G_2$ can be written as powers of $\hat G_1 \cdot \hat G_2$. One can also
restrict the indices on powers of $\hat G_1$ and $\hat G_2$ to be completely
symmetrized, since terms antisymmetric in the indices can be eliminated using
the operator identities. The series in $\hat G_1 \cdot \hat G_2$ terminates
after $N_c$ terms, because an operator with more than $N_c$ quark fields acting
on a single baryon can be reexpressed in terms of operators with $\le N_c$
quark fields.\foot{An easy way to see this is to normal order the operators. A
normal ordered product involving with more than $N_c$ quark operators on a
baryon vanishes, which gives the desired identity.}

There are no $1/N_c$ corrections to \vcentral, except through $1/N_c$
dependence in the unknown coefficients $v_n$. All terms in the $1/N_c$
correction to \vcentral\ are arbitrary polynomials in $G^{ia}_{1,2}$, with one
factor of $S^i_{1,2}$ or $I^a_{1,2}$. It is easy to check that all such terms
have the wrong time-reversal properties to contribute to the baryon-baryon
potential. Thus the first correction to \vcentral\ contains a factor of $\hat
S_1\cdot \hat S_2$ or $\hat I_1\cdot \hat I_2$ and is of order $1/N_c^2$.

Eqs. \hartree, \ijrule, and \genpot\ define the $N_c$-counting for the central
potential.  Large $N_c$ QCD implies that the central potential is of order
$N_c$, but is determined by only two independent functions instead of four (at
leading order in $1/N_c$):
\eqn\vlzero{ V^0_0(r)\sim N_c\ ,\qquad V_\sigma^1(r) \sim N_c\ ,}
while
\eqn\vlzerodif{V^0_\sigma(r)\sim 1/N_c\,\qquad
V_0^1(r)\sim 1/ N_c\ .}
As was noted in \ks\ and will be discussed in \S5, the above relation implies
that the central potential obeys an effective Wigner supermultiplet symmetry.

\subsec{$\Delta L=1$: The Spin-Orbit Potential}

The $\Delta L=1$ baryon interaction amplitude contains the spin-orbit coupling
term; it is obtained from the general Hartree Hamiltonian eq.~\hartree\ by
restricting attention to terms for which  the coefficient $v$ transforms as a
vector under rotations. It follows that the 1-quark operators multiplying the
$v$ coefficients must be combined to transform as a $\(1,0\)$ representation
under $SU(2)_{spin}\times SU(2)_{isospin}$. From eq.~\nsize\ we have seen that
to contribute at leading order in $1/N_c$, an $n$-quark operator must be a
polynomial in the $\hat G$'s alone. However, one cannot make a $\(1,0\)$
operator with the correct parity and time reversal properties purely out of
$\hat G$'s. The spin-orbit force is suppressed relative to the central force,
and is an arbitrary polynomial in $\hat G$'s, with one factor of $\hat S$ or
$\hat I$. The general form of the $\Delta L=1$ amplitude is
\eqn\jone{\openup 1\jot\eqalign{
V_{LS}  =
& N_c \sum_{n=0}^{N_c-1} v_{1,n}^i \left({\hat S_1^i + \hat S_2^i \over N_c}
\right)
 \left({\hat G_1 \cdot \hat G_2 \over N_c^2} \right)^n \cr
 +& N_c \sum_{n=0}^{N_c-2} v_{2,n}^i \left({\hat G_2^{ia} \hat I_1^a + \hat
G_1^{ia}
\hat I_2^a \over N_c^2}\right)
 \left({ \hat G_1 \cdot \hat G_2 \over N_c^2} \right)^n \cr
 +& N_c\sum_{n=0}^{N_c-3}  v_{3,n}^i  \left({ \hat G_1^{ia} \hat G_2^{ja} \hat
S_1^j + \hat
G_2^{ia} \hat G_1^{ja} \hat S_2^j \over N_c^3} \right)
 \left({ \hat G_1 \cdot \hat G_2 \over N_c^2}\right)^n .
}}
This can be derived by arguments similar to those in the previous subsection.
Time reversal and parity invariance requires the coefficients $v_{m,n}^i$ in
\jone\ to be proportional to $\( \vec q \times \vec k\)$ times an arbitrary
function of $\vec q^2$ and $\vec k^2$. In position space, a contribution of the
form
$U(\vec q^2) \( \vec q \times \vec k\)\cdot \(\vec S_1 + \vec S_2\)$ is of the
form $\left(\vec \nabla U(r) \times \vec k \right) \cdot \vec S$,  which is the
usual
spin-orbit force. There is a hidden suppression factor of  $1/N_c$ (which
follows from eq.~\kqsupp) in the spin-orbit force which is not manifest in
eq.~\jone,  since the $\Delta L=1$ interaction necessarily involves both $\vec
q$ and $\vec k$.

The Wigner-Eckart theorem implies that there are only two distinct operators
when the expression \jone\ for the $\Delta L=1$ amplitude  is restricted to
the  nucleon sector. These are the two spin-orbit terms appearing in
eq.~\genpot. Thus we find
\eqn\vlone{ V_{LS}^0(r) \sim 1/N_c\ ,\qquad V_{LS}^1(r)\sim
1/N_c\ .}
The spin-orbit force is $\CO(1/N_c^2)$ in strength relative to the central
force, and it is of comparable strength in the two isospin channels.

\subsec{$\Delta L=2$: The Tensor and Quadratic Spin-Orbit Potentials}

The $\Delta L=2$ amplitude is obtained by requiring that the coefficients
$v$ in eq.~\hartree\ transform under rotations as $\Delta L=2$. The
leading order amplitude is a polynomial in the $\hat G$'s that transforms as
$S=2,\ I=0$. One can obtain an amplitude that does not violate the $I_t=J_t$
rule on each baryon line by combining $I=S=1$ amplitudes on each baryon into
total $I=0$ and total $\Delta L=2$. The general form of the leading order
amplitude is
\eqn\jtwo{
V_T^1 = N_c\sum_{n=0}^{N_c-1} v_n^{ij}
{\hat G_1^{ia} \hat G_2^{ja}\over N_c^2} \left( {\hat G_1 \cdot
\hat G_2 \over N_c^2}\right)^n
}
where the coefficient $v_n^{ij}$ is a symmetric traceless tensor that depends
on $\vec q$ and $\vec k$. Time reversal invariance requires the coefficients to
have the form
$$
v_n\times \( \vec q^i \vec q^j - {1\over3}\vec q^2 \delta^{ij}\ \ {\rm or}\ \
\vec k^i\vec k^j - {1\over3}\vec k^2
\delta^{ij}\),
$$
where $v_n$ is a scalar function of $\vec q^2$, $\vec k^2$ and $\left( \vec q
\cdot \vec k \right)^2$.

If one restricts the interaction eq.~\jtwo\ to the nucleon sector, one gets
\eqn\jtwoii{
V_T^1 = N_c\, v_n\, \tau_1 \cdot \tau_2 \left( q \cdot \sigma_1
\ q \cdot \sigma_2 - {1\over3} q^2 \ \sigma_1 \cdot \sigma_2 \right).
}
Terms with $n>1$ in eq.~\jtwo\ can be dropped, because two spin-1/2 nucleons
can only give non-zero matrix elements for operators with spin $\le 1$.
Comparing with eq.~\genpot, we see that
\eqn\jtwoiii{
V_T^1\sim N_c.
}
The other term in the tensor potential, $V_T^0$, has $\vert I-S\vert=1$ at each
nucleon line, and so by eq.~\ijrule
\eqn\jtwoiii{
V_T \sim 1/N_c.
}
A similar and  straightforward analysis for $V_Q$ gives the
results listed in Table~1.

\subsec{The $NN$ potential and the $\Delta$}

One flaw in our discussion that should be eventually improved upon is the
treatment of the $\Delta$. The Hartree Hamiltonian \hartree\ implicitly acts
on the entire $I=S$ baryon tower, including both nucleons and $\Delta$'s, all
of which
are degenerate in the $N_c\to\infty$ limit. In our discussion of the $NN$
potential, we have simply projected $H$ to the nucleon sector. A more
sophisticated treatment would be to integrate the $\Delta$'s out of the theory
(keeping track of the $1/N_c$ mass splitting) and to
construct  an effective theory for nucleons alone. This is a subtle analysis
(see, for example, \ref\sav{M. Savage,
{\tt nucl-th/961102}}) and beyond the scope of this paper.

\newsec{Comparison of large $N_c$ QCD with a phenomenologically successful
model}

Our large-$N_c$  results for the general nucleon-nucleon potential of
eq.~\genpot\ are displayed in Table~1.  For two flavors we have found that the
strongest $NN$ interactions  are the central force terms $V_0$ and
$V_\sigma^\tau$, as well as the tensor force $V_T^\tau$, all three of which are
$\sim N_c$.  The remaining contributions to the $NN$ potential, with the
exception of $V_Q$, are relatively suppressed by $\sim \CO(1/N_c^2)$. Finally,
the isospin invariant quadratic spin-orbit force $V_Q$ is suppressed by
$\sim\CO(1/N_c^4)$ compared to the central potential, as it is both an $I\ne S$
interaction, as well as being a second order relativistic effect suppressed by
$1/M^2$. The results we have derived are consistent with the $I_t=J_t$ rule,
but are more general. They are true in QCD in the $1/N_c$ expansion, and make
no assumptions about the origin of the $NN$ interaction as being, for example,
due to one meson exchange.

The results can be directly compared with nuclear potential models in momentum
space. A particularly simple phenomenological model to compare with is the
meson exchange model ``Nijmegen potential'' of references \refs{\niji,\nijii}.
In this model,  the $NN$ potential is approximated in momentum space by a sum
of Yukawa and Gaussian interactions times powers of momenta divided by masses,
contracted with the spin and isospin Pauli matrices. The Yukawa potentials
correspond to one-particle exchange of both real mesons($\pi, \eta, \eta',
\rho, \omega, \phi, a_0, f_0$) and an ``effective meson'' ($\epsilon$), while
the Gaussian potentials are labelled by $P$, $f_2$, $f_2'$ and  $a_2$. The
motivation for this form of the potential is unimportant here; it provides a
phenomenologically successful parametrization for the $NN$ potential that can
be compared with Table~1.  The $N_c$ dependence should appear in the relative
strengths of the potentials, and the $1/M$ factors that appear when the
potential is decomposed as in eq.~\genpot.  The strength of the contributions
to the Nijmegen potential are simple to evaluate, since they are presented
explicitly in momentum space, and we can treat all momenta and meson masses as
$\sim 1$ in the $1/N_c$ expansion.  The $N_c$ dependence must then reside in
the strengths of the couplings used in the Nijmegen potential, as well as the
explicit factors of the nucleon mass that appear in the formulas of
ref.~\niji.  One finds for the strength of the various terms in the potential
\eqn\nijn{\openup1\jot\eqalign{
V^I_0&\sim g_{I0}^2,\ {g_{I0} g_{I1}\Lambda \over M},\ {g_{I1}^2\Lambda^2
\over M^2},\cr
V^I_\sigma&\sim V^I_T\sim {g_{I0}^2\over M^2},\ {g_{I0}g_{I1} \over \Lambda M},
\ { g_{I1}^2 \over \Lambda^2},\cr
V^I_{LS}&\sim {g_{I0}^2\over M^2},\ {g_{I0}g_{I1} \over \Lambda M},\cr
V^I_Q&\sim {g_{I0}^2\over M^4},\ {g_{I0} g_{I1} \over \Lambda M^3},
\ { g_{I1}^2 \over \Lambda^2 M^2}
\ ,}}
where $I=0,1$ correspond  to the $1\cdot 1$ and $\tau_1\cdot \tau_2$ isospin
structures respectively, $M$ is the nucleon mass, and $\Lambda$ is a strong
interaction scale characterizing the derivative expansion (denoted $\CM$ in
\niji).  The parameters $g_{IS}$ correspond to  the coupling constants of the
model with  $t$-channel
(isospin, spin)$=(I,S)$ in the nonrelativistic limit; in particular, the scalar
coupling $g_S$ and vector couplings  $g_V$ and $f_V$ of ref.~\niji\ are given
by $g_{I0}$, $g_{I0}$
and $g_{I1}$ respectively, where $I$ is the meson isospin.  (The pseudoscalar
contributions are parametrized differently in \niji\ and are mentioned below).
As far as the $N_c$
scaling goes, $M\sim N_c$, while the $\Lambda$ and the masses of the
exchange mesons are all $\sim 1$. In eq. \nijn\  we have omitted dimensionful
quantities that do not scale with
$N_c$, such as the meson propagators $1/(q^2+m^2)$.  By comparing the
expressions in eq. \nijn\ with our results in Table~1, one sees that they are
consistent provided that the couplings
$g_{IS}$  scale with $N_c$ as
\eqn\gfit{g_{IS}\propto N_c^{(1/2 -|I-S|)}\ .}
This $N_c$ scaling can be compared with the numerical values given in
ref.~\nijii. In fig.~3 we have plotted the couplings determined numerically in
ref.~\nijii, rescaled by their value for $f_\rho$.  Since $f_\rho$ is a
$g_{11}$ coupling, eq. \gfit\ implies that the leading large-$N_c$ prediction
for the ratio is
\eqn\ratpred{ {\hat g}_{IS} \equiv {g_{IS}\over f_\rho} = \cases{1, &if
$|I-S|=0$\cr \frac{1}{3}, &if $|I-S|=1$\cr}\ .}

As can be seen from fig.~3, there is good qualitative agreement between the
large-$N_c$ prediction \ratpred, and the $g_{IS}$ values used in the Nijmegen
potential. Omitted from fig.~3 are the pseudoscalar couplings, which are not
readily compared with heavy meson couplings, due to their special status as
pseudo-Goldstone bosons. However, the pseudoscalar meson couplings are related
to the axial current couplings, which have been analyzed in detail, and shown
to agree with $1/N_c$ predictions \ref\dai{J.~Dai, R.~Dashen, E.~Jenkins, and
A.V.~Manohar, \physrev{D53}{1996}{273}.}.  There are two couplings in the
Nijmegen potential, the $\phi$ and $a_2$ coupling, that deviate significantly
from the
$1/N_c$ pattern. The $\phi$ meson is a pure $\bar s s$ state, and only couples
to the nucleons through quark loops. Its coupling is OZI suppressed, and should
be of order $1/N_c$ relative to the $\omega$ couplings. The Nijmegen fit has
$g_\phi/g_\omega \approx 0.1$, which is a factor of three smaller than the
naive $1/N_c$ prediction. The $a_2$ coupling is even somewhat smaller.

\topinsert
\centerline{\epsfxsize=3.7in \epsfbox{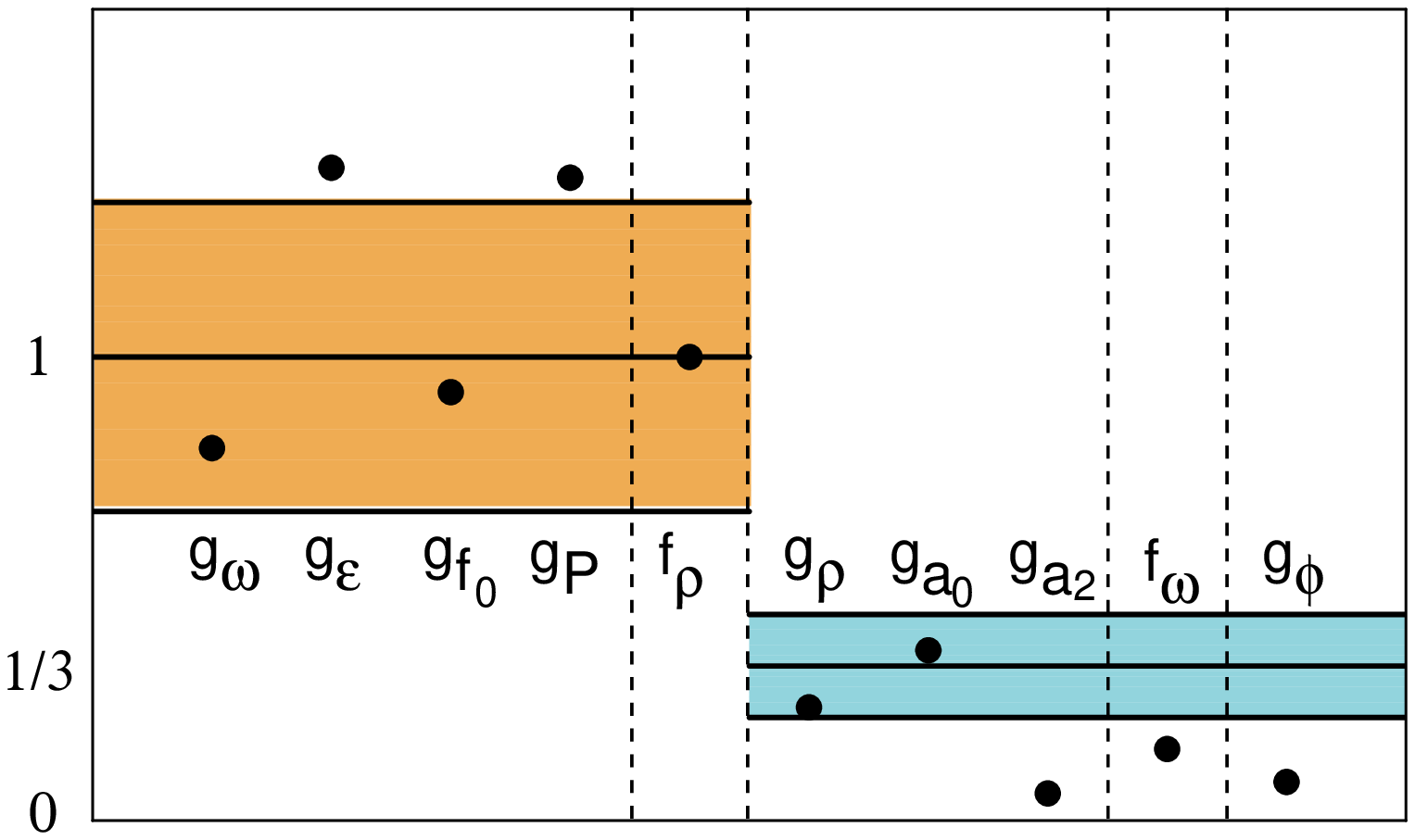}}
\caption{Fig.~3. The couplings for the Nijmegen $NN$ potential in ref.~\nijii\
rescaled by $f_\rho$.  The values for this ratio predicted by large $N_c$ QCD
in eq.~\ratpred\ are indicated by lines, and the shaded regions are the size of
the expected $\CO(1/N_c)$ corrections to the leading result. The five regions
in the plot (separated by vertical dashed lines) are (from left to right) the
$(I,S)=(0,0)$,
$(1,1)$, $(1,0)$, $(0,1)$, and  $g_\phi$ couplings.}
\endinsert
\nfig\figiv{The couplings for the Nijmegen $NN$ potential in ref.~\nijii\
rescaled by $f_\rho$.  The values for this ratio predicted by large $N_c$ QCD
in eq.~\ratpred\ are indicated by lines, and the shaded regions are the size of
the expected $\CO(1/N_c)$ corrections to the leading result. The five regions
in the plot (separated by vertical dashed lines) are (from left to right) the
$(I,S)=(0,0)$,
$(1,1)$, $(1,0)$, $(0,1)$, and  $g_\phi$ couplings. }

It must be stressed that the numerical parameters plotted in fig.~3 were
obtained by treating the couplings as phenomenological parameters in the $NN$
potential, chosen to provide the best fit to $NN$ scattering data. There is no
reason to assume that the $NN$ force is actually due to single meson exchange;
in fact, the $\epsilon$ and $P$ contributions to the potential do not
correspond to single meson exchange at all, and the $a_2$ in the Nijmegen
potential has a Gaussian propagator. The model subsumes such effects as 2-$\pi$
exchange, $\rho\pi$ exchange, etc. within the phenomenological couplings
$g_{IS}$.  Only the pseudoscalar meson couplings are related to the physical
meson-nucleon couplings, since the long distance part of the $NN$ potential is
dominated by single meson exchange. Thus the agreement between fig.~3 and the
large-$N_c$ prediction \ratpred\ contains more than the claim that meson-baryon
couplings obey the $I_t=J_t$ rule. We take fig.~3 to provide encouraging
evidence that
our large-$N_c$ analysis of the $NN$ interaction describes the qualitative
features seen in nature.

\newsec{The Central Potential and Wigner Supermultiplet Symmetry}

It was suggested in ref. \ks\ that the approximate Wigner supermultiplet
symmetry observed in light nuclei could be explained by the $1/N_c$ expansion
of QCD.  Under the Wigner symmetry $SU(4)_W$, the four {\it nucleon} states
$p\uparrow$,
$p\downarrow$, $n\uparrow$ and $n\downarrow$ transform as the four dimensional
fundamental representation. Note that $SU(4)_W$ is distinct from the quark
model $SU(4)$, and that the former cannot be realized as a symmetry at the
quark level. Nevertheless, ref. \ks\  argued that the $1/N_c$ expansion
explains how $SU(4)_W$ symmetry could emerge as an accidental symmetry in light
nuclei.   As  that work only examined the central part of the $NN$ potential,
it is worth reexamining the argument.

Under $SU(4)_W$ symmetry, a two-nucleon state transforms like
${\bf 4}\times {\bf 4} = {\bf 6}_A + {\bf 10}_S$, where the
subscripts $A$ and $S$ denote the antisymmetric and symmetric
combinations. Under ${\rm spin \times isospin}$, these representations
decompose as
\eqn\sufourdec{\eqalign{
{\bf 6} &\rightarrow \left(0,1\right)+\left(1,0\right),\cr
{\bf 10} &\rightarrow \left(0,0\right)+\left(1,1\right)\ .
}}
If the two nucleons are in an
even partial wave,  they must be in a totally antisymmetric spin $\otimes$
isospin state, so they are in a $\left(0,1\right)$ or $\left(1,0\right)$ state,
i.e. in ${\bf 6}_A$ of $SU(4)_W$.  If the two nucleons are in an odd partial
wave, they must be in a totally symmetric spin $\otimes$ isospin state, so they
are in a $\left(0,0\right)$ or $\left(1,1\right)$ state, i.e. in ${\bf 10}_S$
of $SU(4)_W$.

We have shown in \S3 that the leading contributions to the $NN$ potential (at
strength $N_c$) are $V_0^0$, $V_\sigma^1$ and $V_T^1$.  The first two terms
correspond to the operators
\eqn\wigop{
1,\qquad \sigma_1 \cdot \sigma_2\ \tau_1 \cdot \tau_2, \qquad
}
which have the same value on $\left(0,1\right)$ or $\left(1,0\right)$, i.e.
they have
the same value on the entire ${\bf 6}$ representation of $SU(4)_W$. Thus at
leading order in $N_c$, the
central potential respects Wigner $SU(4)_W$ symmetry if the two nucleons are in
an even partial wave. Violation of Wigner $SU(4)_W$ from the central potential
in the even partial waves
is an $\CO(1/N_c^2)$ effect. The operators
eq.~\wigop\ have different values on the $\left(0,0\right)$ and
$\left(1,1\right)$ representations, and so break $SU(4)_W$ symmetry when acting
on the ${\bf 10}$ representation of $SU(4)_W$. Thus the central potential
breaks Wigner $SU(4)_W$ symmetry  at leading order in the odd partial waves.
The tensor force $V_T^\tau$ also violates Wigner $SU(4)_W$ symmetry at leading
order, in all partial waves.

Nevertheless, there is reason to expect to see Wigner symmetry in light nuclei.
The nucleons inside a nucleus have low momentum, so the dominant
interaction is $s$-wave scattering, with higher partial waves being
kinematically suppressed.  Furthermore, the tensor mean field is small in
nuclei.  Therefore all of the leading order violations of $SU(4)_W$ may be
expected to be small.

So why is $SU(4)_W$ not evident in heavy nuclei?
At subleading order (a relative $1/N_c^2$), the potentials $V_\sigma^0$,
$V_0^1$, $V_T^0$ and $V_{LS}^{0,1}$ all break the Wigner symmetry.  The mean
fields of all but the spin-orbit force are small in nuclei.  However, the
importance of the spin-orbit force grows like $A^{1/3}$, proportional to the
number of particles in the maximum angular momentum shell.  Therefore, for
large $A$,  the spin-orbit force is expected to  overcome the $1/N_c^2$
suppression and destroy the approximate Wigner supermultiplet symmetry. It may
be interesting to pursue this further, to determine at what values of $A$ one
might expect $SU(4)_W$ symmetry to fail.

\newsec{Summary and Conclusions}

The $1/N_c$ expansion has been shown elsewhere to be a useful
tool in analyzing the properties of baryons \foot{See for example,
\nref\jl{E.~Jenkins and R.F.~Lebed, \physrev{D52}{1995}{282}}\nref\panic{A.V.
Manohar,  hep-ph/9607484, Talk given at 14th International Conference on
Particles
and Nuclei (PANIC 96),  Williamsburg, VA, 22-28 May 1996.}\refs{\jl,\panic}.};
the analysis presented here and in ref. \ks\ shows
that it also provides a useful tool for understanding qualitative features of
the nuclear force.  In particular, we have computed the relative strengths of
the various components of the $NN$ interaction in the $1/N_c$ expansion
(Table~1) and argued that the predicted patterns are reproduced in
phenomenological models of the $NN$ force (fig.~3).  We also extended the
argument of ref. \ks\ that the approximate Wigner supermultiplet symmetry
observed in light nuclei (see \ks\ for examples and references) is in fact
understandable in terms of the $1/N_c$ expansion.  We are aware of no other
explanations for this peculiar $SU(4)_W$ symmetry.

Aside from obtaining directly from QCD a qualitative explanation for the spin,
isospin and tensor structure of the $NN$ potential, it is hoped that the
$1/N_c$ expansion could serve as a guide toward better understanding the
interactions of baryons with strangeness, where the experimental data is much
poorer.  To this end we have included the three flavor analysis in  Appendix~A.
Our hope is that this could prove useful for understanding  hypernuclei, as
well as matter in extreme conditions where strangeness may play a significant
role, such as in heavy ion collisions, or dense matter with kaon condensation
\ref\kcon{D.B. Kaplan and A.E. Nelson, \pl{175}{1986}{57}, B179 (1986) 409E;
\pl{192}{1987}{193}.}\ or hyperons.

\newsec{Acknowledgments}

We would like to thank G. Bertsch J.-L. Forest, V. Pandharipande
and M.J. Savage for useful conversations.
D.K. was supported in part by DOE grant DOE-ER-40561, and
NSF Presidential Young Investigator award \pyidk.
A.M. was supported in part by Department of Energy grant
DOE-FG03-90ER40546.

\listrefs

\appendix{A}{Three Flavors}

The quark operator basis for three flavors is denoted by
\eqn\Aquarkbasis{
\hat \CS^i = Q^\dagger {\sigma^i\over 2} Q,\qquad \hat \CT^A =
Q^\dagger T^A Q,\qquad \hat
\CG^{iA} =
Q^\dagger {\sigma^i \over 2} T^A Q,
}
where $Q=(u,d,s)$ and $Q^\dagger$ are the creation and annihilation operators
for the three light quark flavors, $i=1,2,3$ and $A=1,\ldots,8$. $T^a$ are the
standard $SU(3)$ matrices in the fundamental representation, normalized so that
$\tr T^A T^B=\delta^{AB}/2$. These 1-quark operators act on a baryon state
which is the completely symmetric tensor product (in spin $\otimes$ flavor) of
$N_c$ quarks.

It is convenient to break the operator basis \Aquarkbasis\ for the $SU(6)$
generators by separating $Q=(u,d,s)$ into $q=(u,d)$ and $s$. Under this
decomposition $\hat \CS^i$, $\hat \CT^A$ and $\hat \CG^{ia}$ break up into
linear combinations of
\eqn\Anewbasis{\eqalign{
&\hat S^i = q^\dagger {\sigma^i \over 2} q,
\qquad \hat I^a = q^\dagger {\tau^a\over 2} q,
\qquad \hat G^{ia} = q^\dagger {\sigma^i \tau^a \over 4} q, \cr
&\hat S_s^i = s^\dagger {\sigma^i \over 2} s,
\qquad \hat N_s = s^\dagger s, \qquad \hat Y^{i\alpha} = s^\dagger {\sigma^i
\over 2} q^\alpha,
\qquad \hat K^\alpha = s^\dagger q^\alpha,
}}
and $\hat Y^{i\alpha\dagger}$ and $\hat K^{\alpha\dagger}$ which are the
hermitian conjugates of $Y^{i\alpha}$ and $K^\alpha$. For baryons with $N_c$
quarks, and strangeness of order one, $\hat G^{ia}$ is of order $N_c$, $\hat
Y^{i \alpha}$ and $\hat K ^\alpha$ are of order $\sqrt{N_c}$, and $\hat
S^i_{ud}$, $\hat I^a$, $\hat S_s^i$ and $\hat N_s$ are of order one \djm. Note
that $\hat Y^{i\alpha}$ and $\hat K ^\alpha$ are strangeness changing
operators.

The (properly normalized) $SU(6)$ generators $\sqrt2\hat \CG^{iA}$, $\hat
\CT^A/\sqrt2$ and $\hat \CS^i/\sqrt 3$ are collectively denoted by $\hat
\Lambda^M$. The operator basis for two and three flavors are summarized in
Table~2. An expansion using the operator basis \Anewbasis\ gives us the
predictions of the $1/N_c$ expansion for three flavors, without assuming
$SU(3)$ symmetry. One can also impose $SU(3)$ symmetry, which places additional
restrictions on the final result. The results for two flavors are obtained by
using only the operators $S^i$, $I^a$, and $G^{ia}$.
\midinsert
\centerline{TABLE 2}
\centerline{\vbox{
\medskip
\tabskip=0pt \offinterlineskip
\def\space#1{height #1 pt&\omit&&\omit&&\omit&&\omit&&\omit&&\omit&\cr}
\def\space#1{height #1 pt&\omit&&\omit&&\omit&&\omit&&\omit&&\omit&\cr}
\halign{\vrule # &\quad\hfil # \hfil\quad&& \vrule # &
\quad\hfil $ # $ \hfil\quad \cr
\noalign{\hrule}
\space 3
& {\rm \# of Flavors} && && {\rm Spin} && {\rm Flavor} && {\rm Spin-Flavor}
 && {\rm All } &\cr
\space 3
\noalign{\hrule}
\space 3
& 2 && && \hat S^i && \hat I^a && \hat G^{ia} && \hat \lambda^\mu &\cr
\space 3
\space 3
& 3 && && \hat \CS^i && \hat \CT^A && \hat \CG^{iA} && \hat \Lambda^M &\cr
\space 3
\space 3
&  && \Delta S =0 \hfill && \hat S^i && \hat I^a, \hat N_s && \hat G^{ia},
\hat S_s^i && &\cr
\space 1
& $ 3\rightarrow 2$ && \Delta S = 1 \hfill && && \hat K^\alpha && \hat
Y^{i\alpha} &&
&\cr
\space 1
&  && \Delta S = -1 \hfill  && && \hat K^\dagger_\alpha && \hat
Y^{i\dagger}_\alpha
&& &\cr
\space 3
\noalign{\hrule}
}}
}
\endinsert

The results of the paper can be generalized to the case of three
light flavors. The analysis is more complicated because one also has to
include operators $\hat N_s$, $\hat S_s^i$, $\hat Y^{i\alpha}$ and $\hat
t^\alpha$ that involve the $s$ quark. We will simply give the final results
here.

The $1/N_c$ $\Delta L=0$ interaction is
\eqn\Aaiv{
{\CA^{j=0}_1 \over N_c}=
\sum_{r=0}^{N_c}
 c_{1,r} \left({\hat \Lambda_1\cdot \hat \Lambda_2\over N_c^2}\right)^r
+\sum_{r=0}^{N_c-1} c_{2,r} \epsilon {\hat N_{s1} + \hat N_{s2}\over N_c}
\left( { \hat \Lambda_1 \cdot
\hat \Lambda_2 \over N_c^2} \right)^r.
}
 The $N_s$ term violates $SU(3)$
symmetry, so its coefficient is proportional to $SU(3)$ breaking in the baryon
sector, which is parameterized by $\epsilon$, a dimensionless number of order
0.3. It is clear from Eq.~\Aaiv\ that the $N_s$ term violates $SU(6)$ symmetry
but respects $SU(4)$ symmetry, so that $SU(6)$ violation is of order
$\epsilon/N_c$, but $SU(4)$ violation is of order $1/N_c^2$.

The $\Delta L=1$ interaction for three flavors is
\eqn\Ajoneiii{\eqalign{
{\CA^{j=1} \over N_c^2} =
& \sum_{r=0}^{N_c-1} d_{1,r}^i \left({\hat \CS_1^i + \hat \CS_2^i \over N_c}
\right)
 \left({\hat \Lambda_1 \cdot \hat \Lambda_2 \over N_c} \right)^r \cr
& + \sum_{r=0}^{N_c-2} d_{2,r}^i  \left({\hat \CG_2^{iA} \hat T_1^A + \hat
 \CG_1^{iA}
\hat T_2^a \over N_c^2}\right)
 \left({ \hat \Lambda_1 \cdot \hat \Lambda_2 \over N_c^2} \right)^r \cr
& +\sum_{r=0}^{N_c-3}  d_{3,r}^i  \left({ \hat \CG_1^{iA} \hat \CG_2^{jA}
\hat \CS_1^j + \hat
\CG_2^{iA} \hat \CG_1^{jA} \hat \CS_2^j \over N_c^3} \right)
 \left({ \hat \Lambda_1 \cdot \hat \Lambda_2 \over N_c^2}\right)^r \cr
 &+\sum_{r=0}^{N_c-1} d_{4,r}^i \epsilon
 \left( { \hat S_{1s}^i + \hat S_{2s}^i
 \over N_c} \right)
 \left({ \hat \Lambda_1 \cdot \hat \Lambda_2 \over N_c^2} \right)^r \cr
 & + \sum_{r=0}^{N_c-3} d_{5,r}^i \epsilon \left({
 \hat \CG_1^{iA} \hat \CG_2^{jA}
\hat  S_{1s}^j + \hat \CG_2^{iA} \hat \CG_1^{jA} \hat S_{2s}^j\over N_c^3}
\right)
 \left({\hat \Lambda_1 \cdot \hat \Lambda_2 \over N_c^2} \right)^r,
}}
which is the three-flavor generalization of \jone. Time reversal and parity
invariance requires the coefficients in \jone\ to be of the form
$$
\pin \times \pout,
$$
times an arbitrary function of $q^2$, $k^2$ and $\left( q \cdot k \right)^2$.
As for the case of two flavors, the coefficients in eq.~\Ajoneiii\ are of order
$1/N_c$, so that the $\Delta L=1$ amplitude is of order $1/N_c^2$ relative to
the central potential.

The $\Delta L=2$ amplitude is
\eqn\Ajtwoii{
{\CA_0^{j=2}\over N_c} = \sum_{r=0}^{N_c-2} b_r^{ij}
{\hat \CG_1^{iA} \hat \CG_2^{jA}\over N_c^2} \left( {\hat \Lambda_1 \cdot
\hat \Lambda_2 \over N_c^2}\right)^r
}
where the coefficient $f_r^{ij}$ is a symmetric traceless tensor that
depends on $\pin$ and $\pout$. This is the three-flavor generalization of
\jtwo.

\bye